\documentstyle[preprint,aps,prc,array,epsf]{revtex}
\input{psfig}

\newcommand{\Del}{$\Delta$}

\def\beq{\begin{equation}}
\def\eeq{\end{equation}}
\def\bea{\begin{eqnarray}}
\def\eea{\end{eqnarray}}
\def\eps{\varepsilon}    
\def\eqref#1{Eq.~(\ref{eq:#1})}
\def\eqlab#1{\label{eq:#1}}
\def\figref#1{Fig.~(\ref{fig:#1})}
\def\figlab#1{\label{fig:#1}}

\def\VYP#1#2#3{{\bf #1} (#2) #3}  
\def\NP#1#2#3{Nucl. Phys. {\bf #1} (#2) #3}
\def\PL#1#2#3{Phys.~Lett. {\bf #1} (#2) #3}
\def\PR#1#2#3{Phys.~Rev.~{\bf #1} (#2) #3}
\def\PRL#1#2#3{Phys.~Rev.~Lett. {\bf #1} (#2) #3}

\newcommand{\thalf}{\mbox{\small{$\frac{3}{2}$}} }
\def\half{\mbox{\small{$\frac{1}{2}$}}}

\begin{document}
\draft
\tighten
\title {$Q^2$-Dependence of the Drell-Hearn-Gerasimov integral}
\author{O. Scholten$^{a}$ and A.Yu. Korchin$^{b,}$\footnote{Permanent
address: National Science Center ``Kharkov Institute of Physics and
Technology'', 310108 Kharkov, Ukraine}
}
\address
{$^a$ Kernfysisch Versneller Instituut, Zernikelaan 25, 9747 AA Groningen, The
Netherlands}
\address
{$^b$ Department of Subatomic and Radiation Physics, University of
Gent, Proeftuinstraat 86, B-9000 Gent, Belgium}

\maketitle

\begin{abstract}

The energy and four-momentum ($Q^2$) dependence of the photo-absorption cross
section on the proton is calculated for helicity -\half\ and -\thalf states.
An effective Lagrangian model is used, formulated in terms of meson and baryon
degrees of freedom, which obeys crossing symmetry, unitarity, Lorentz and gauge
invariance. The difference in the cross sections for the two helicity states,
the Drell-Hearn-Gerasimov integral $I_{DHG}(Q^2)$, is evaluated at
different $Q^2$.
We find that at small momentum transfer the absolute value of $I_{DHG}(Q^2)$
first increases to reach a maximum at $Q^2 \approx 0.05$ GeV$^2$ before
decreasing at higher $Q^2$.

\end{abstract}

\vspace{2cm}

\bigskip


\section{Introduction}

   Absorption of virtual photons on the nucleon at very high
photon four-momentum
($Q^2$) has been proposed as a means to measure the spin content of the
nucleon. Since helicity is conserved in the scaling limit, the cross-section
difference between parallel and anti-parallel helicities for the photon and
the proton should be a measure of the spin carried by the quarks in the
proton. This difference, integrated over energy, is called the DHG integral.
At large $Q^2$ this integral has been expressed in terms of a sumrule
by Ellis and Jaffe
\cite{Ell74} (EJ). Data at high $Q^2$ seem to agree with the momentum
dependence predicted by this QCD-based sumrule. The magnitude is however
considerably smaller than predicted and this discrepancy has been known as
the ``spin crisis". For real photons, $Q^2=0$, another, rigorous sumrule
has been formulated for this integral by Drell and Hearn \cite{Dre66} and
independently by
Gerasimov \cite{Ger66} (DHG). While the values predicted by EJ are positive,
the DHG value is large and negative and it is an intriguing problem to
reconcile the two.

The different values for the sumrule at low and high $Q^2$ are
related to the transition from physics dominated by nucleon resonances
\cite{Clo72} (non-perturbative QCD) to the perturbative QCD regime.
Several studies \cite{Bur92,Bur93,Dre95,Ma97} have
emphasized the explicit role played by nucleon resonances in the transitional
regime around $Q^2 \approx 1$ GeV$^2$. At the
lowest momentum transfers this has been investigated in
chiral perturbation theory \cite{Ber93}.

The derivation of the sumrules is based on Lorentz and gauge invariance,
crossing symmetry, unitarity and causality. It is therefore of interest to
investigate the sumrule in a model in which most of these symmetries are
obeyed. We present a calculation of the strength distribution for $Q^2 \leq
1$ GeV$^2$ in the model developed in ref. \cite{Pas95,Kor98b}. This model
obeys crossing symmetry, unitarity, Lorentz and gauge invariance. It is
formulated in terms of meson and nucleon degrees of freedom which includes
nucleon resonances in an effective-Lagrangian formalism.

\section{Outline of the model}

We use the relativistic effective-Lagrangian
formalism presented in \cite{Pas95,Kor98b}. The model is based on the
K-matrix approach. The kernel is constructed from the
 direct (s), exchange (u) and meson exchange (t-channel) tree-level
amplitudes. In the s- and u-channels all spin -1/2 and -3/2
baryon resonances with masses below 1.7 GeV are included. The use of the
K-matrix approach
guarantees unitarity in the coupled-channel $(\gamma + N) \otimes (\pi +N)$
space.
Observing unitarity is of crucial importance for the calculation of cross
sections for photon energies exceeding 250 MeV. Coupling to channels outside
this model space is included in an approximate manner
through the introduction of an imaginary part in the self-energy of the
s-channel resonances \cite{Kor98b}. The coupling parameters have been obtained
from a simultaneous fit to pion-nucleon phase shifts, pion-photoproduction
multipoles and cross sections for Compton scattering \cite{Kor98b}.
The model is Lorentz and gauge invariant and obeys crossing symmetry.
The chiral-symmetry constraints are also respected since the
low-energy $\pi N$ scattering is well described. Here we will only mention
the details which are of interest for the present application.

Of particular importance for the DHG integral is the treatment of the
\Del-resonance. The most general $\gamma N \Delta$ vertex for finite
$q^2=-Q^2$ is given by \cite{Kor97,Kor98a}
\bea
\Gamma_{N\gamma \to \Delta^\alpha} &=& {i\over 2M} F_{VMD}(q^2)\, [
G_1\, \theta_{\alpha\beta}(z_1) \,\gamma_\delta
-{G_2\over 2M} \,\theta_{\alpha\beta}(z_2 ) \,p_\delta
-{G_3\over 2M} \,\theta_{\alpha\beta}(z_3 ) \,q_\delta
 ]\gamma_5\,(q^\beta \eps^\delta-q^\delta \eps^\beta) \;\; \eqlab{Del-vtx}
 \\
&&\theta_{\alpha\beta}(a_i ) = g_{\alpha\beta}+a_i \gamma_\alpha
 \gamma_\beta \ \mbox{\ \ and\ \ } a_i \equiv -(z_i +1/2) \ \mbox{\ \ for\ \ }
i=1,2,3\nonumber
\eea
 where $G_i=g_{i}\;T_3 $ and $T_3$ is the $N\leftrightarrow \Delta$ isospin
transition operator.
The constants $g_1$, $g_2$ and off-shell parameters $z_1$, $z_2$ are
fixed from the fit \cite{Kor98b} for
the real photons. The constant $g_3$ (and $z_3$) does not contribute
for $Q^2=0$ and therefore we will choose $g_3 =0$ in the present investigation.
In general, at finite $Q^2$ the coupling $g_3$ affects the longitudinal
multipole $L_{1+}$ as well as the ratio $E_{1+}/M_{1+}$
for the \Del-resonance.
The transition form factor is taken in the form
\beq
F_{VMD}(q^2)
={2 m_\rho^4 \over (2m_\rho^2 - q^2)(m_\rho^2 - q^2)} \;,
\eqlab{VMD}
\eeq
as inspired by vector-meson dominance where $m_\rho$ is the $\rho$-meson mass.

The DHG integral at finite $Q^2$ can be introduced as
  \beq
I_{DHG}(Q^2) = {2M^2 \over Q^2} \int_0^1 dx
\left( g_1(x,Q^2) - {4 x^2 M^2 \over Q^2} g_2(x,Q^2) \right)
       = {M^2 \over 4 \pi^2 \alpha}  \int_{Q^2/2M}^{\infty}
{d\nu \over \nu} \sigma^{TT}
\label{eq:defin}
  \eeq
which relates this sumrule directly to the transverse-transverse
interference cross section
\begin{eqnarray}
\sigma^{TT} = \frac{1}{2}(\sigma^T_{1/2} -\sigma^T_{3/2}) \;
\end{eqnarray}
for inelastic electron scattering on the nucleon.
Throughout this paper we use the Bjorken variable $x=Q^2/2M\nu$ and
$\nu=p\cdot q/M$, the energy of the virtual photon in the lab system.
The total absorption cross section for a
transverse virtual photon in a state with total helicity $\lambda$ is
denoted by $\sigma^T_{\lambda}$ where the dependence on $\nu$ and $Q^2$ is
not indicated for ease of writing.

The  spin-dependent structure functions which enter in \eqref{defin} are
defined as
\begin{eqnarray}
g_1(x,Q^2)&=&{M \nu \over 4 \pi^2 \alpha (1+Q^2/\nu^2)}
\left( \sigma^{TT} + { \sqrt{Q^2} \over \nu}
\sigma^{LT}_{1/2} \right) \;,\\
g_2(x,Q^2)&=&{M \nu \over 4 \pi^2 \alpha (1+Q^2/\nu^2)}
\left( - \sigma^{TT} + {\nu \over \sqrt{Q^2}}
\sigma^{LT}_{1/2} \right) \;,
\end{eqnarray}
where $\sigma^{TL}_{1/2}$ is the transverse-longitudinal
interference cross section, suppressing again the energy and momentum
dependence. Note that structure functions $G_{1,2}$ introduced by
Bjorken \cite{Bjo66} are related to $g_{1,2}(x,Q^2)$ through
\begin{eqnarray*}
 M^2 \nu G_1(\nu,Q^2)  &=&g_1(x,Q^2) \; ,\\
 M \nu^2 G_2(\nu, Q^2) &=& g_2(x, Q^2) \;.
\end{eqnarray*}
 It should be noted that at finite $Q^2$ \eqref{defin} differs from both
\cite{Dre95} and \cite{Ma97}. In particular,
in \cite{Ma97} the DHG integral
in addition to $\sigma^{TT}$ contains also $\sigma^{LT}_{1/2}$ contribution.
Our definitions agree with \cite{Sco97} and have been choosen since they
yield the expression for the DHG integral as measured in recent
experiments.
In the limit of real photon or in the scaling limit, ($Q^2$, $\nu$)
$\rightarrow \infty$ at fixed $x=Q^2/2M\nu$, above differences vanish.

At the photon point \cite{Dre66,Ger66}
\begin{eqnarray}
I_{DHG}(Q^2=0) = -\frac{1}{4}\kappa^2 \;
\end{eqnarray}
with $\kappa$ being the anomalous magnetic moment of the nucleon,
and in the scaling regime
\begin{eqnarray}
I_{DHG}(Q^2)\rightarrow \frac{2M^2}{Q^2}\int_0^1 {g_1 (x) dx}=
\frac{2M^2}{Q^2}\Gamma_1 \;,
\end{eqnarray}
where $\Gamma_1$ is the moment of $g_1$. Experiment gives for the proton
$\Gamma_1^p \approx 0.126$ at $Q^2 =10.7$ GeV$^2$ \cite{EMC} while the
prediction of EJ \cite{Ell74} is $\Gamma_1^p=0.185$.

\section{Results}

The total photo-nucleon cross section can be calculated from
the imaginary part of the forward-scattering $\gamma^* N \rightarrow
\gamma^* N$ amplitude for total
helicity-\thalf and -\half\ states. In \figref{DHG} the
cross sections for the two initial helicity states are plotted versus energy
$\omega=(s-M^2)/2M$ where $\sqrt{s}$ is the invariant energy of the
system. The energy $\omega$ is related to the integration variable
in \eqref{defin} through
\beq
\omega=\nu-Q^2/2M \eqlab{om-nu}
\eeq
and has the advantage that the s-channel resonances occur at a value
of $\omega$, independent of $Q^2$.
The large peak in the cross section at $\omega \approx 300$ MeV is due to the
\Del-resonance while the one at $\omega \approx 700$ MeV is due to the
$D_{13}$-resonance.

At energies below the \Del-resonance the pion-photon seagull term, which is
needed to ensure gauge invariance for the pion-electroproduction amplitude,
gives by
far the dominant contribution. It contributes to the helicity-\half\ states
only and thus it gives a sizable positive contribution
to the DHG integral at $Q^2$=0. Since this contribution to the cross section
is inversely proportional to the momentum of the virtual photon, it
strongly diminishes when $\sqrt{Q^2}\approx \nu$
causing the decrease (increase of the absolute magnitude) of the DHG
integral seen in \figref{DHG-q} at low $Q^2$. The
dominant contribution to the DHG integral originates from the \Del-resonance
and is negative in sign.
Only at values of $Q^2$ of the order of the $\rho$-meson mass the form
factor \eqref{VMD} starts to cut this \Del-contribution giving rise to
a general decrease of the absolute magnitude of the DHG integral seen in
\figref{DHG-q}.  With
increasing $Q^2$ the absolute value of $I_{DHG}(Q^2)$ thus first increases to
reach a maximum at $Q^2$=0.05 GeV$^2$ after which it strongly decreases.

The above features of the DHG integral can also
be observed from \figref{DHG-nu} where the dependence of the DHG integral
on the upper integration limit
\beq
I_{DHG}^{up}(Q^2) = {M^2 \over 4 \pi^2 \alpha}\int_{Q^2/2M}^{\nu_{up}}
{d\nu \over \nu} \sigma^{TT} \;
\eqlab{pDHG}
\eeq
is given as function of $\omega_{up}=\nu_{up}-Q^2/2M$ (see \eqref{om-nu}).
 It can be seen that the important contribution
to this integral is generated in the region of the \Del-resonance. At
$\omega_{max}\approx 800$ MeV, the maximum energy where we can apply the
present model with confidence, about 80\% of the DHG sumrule value (at
$Q^2=0$) is reached.

In \figref{DHG-q} the single-pion production contribution to $I_{DHG}(Q^2)$
is shown also. At small $Q^2$ it contributes 86\% to the
integral, this fraction is somewhat larger than 76\% as was found by
Karliner \cite{Kar73} at the real-photon point. At higher $Q^2$ the 
multiple-pion emission
contribution decreases in absolute magnitude but increases in relative
importance to 23\% at $Q^2$=1 GeV$^2$.

Finally, as we mentioned, the coupling parameter $g_3$ in \eqref{Del-vtx}
has been chosen zero. In general, at finite $Q^2$ the DHG integral
depends on its value. We have checked that
for a moderate positive value, $g_3 \approx |g_1|$, the DHG integral
changes sign and becomes positive at $Q^2 \approx 1$ GeV$^2$.
The effect of this coupling on the DHG integral and pion electro-production
multipoles will be studied in detail in a separate publication.

\section{Conclusions}

Our calculations in an effective-Lagrangian model show that at 
$Q^2 < 1$ GeV$^2$ the DHG integral is dominated
by resonance contributions, mainly by the P$_{33}$ and the 
D$_{13}$-resonances. Due to the form factors which are implemented
in the calculation these contributions decrease rather sharply at moderate
$Q^2$. At large $Q^2$ exceeding a few GeV$^2$ the present model, being based
on nucleon and meson degrees of freedom, looses validity since the quark
structure of the particles starts to play an increasingly important role.
In the framework of an effective-Lagrangian approach one could model the 
quark structure by ascribing different Q$^2$-dependences to electric and 
magnetic couplings of resonances. This might well explain the
change of sign of the integral at large $Q^2$.

At small $Q^2$, $Q^2 < 0.05$ GeV$^2$, we observe a striking increase in
the absolute magnitude of the DHG integral. This is due to the particular
momentum dependence of the seagull (or $\gamma \pi NN$) contribution to the
dominant pion-electroproduction process.
Surprisingly this behavior of $I_{DHG}(Q^2)$ is opposite to the dependence
obtained in the ChPT calculations \cite{Ber93}.

\section*{Acknowledgments}
We would like to thank R.G.E. Timmermans and R. Van de Vyver for discussions.
O.S. gratefully acknowledges the hospitality of RCNP, Osaka, where this work
was initiated. This work is supported by the Fund for
Scientific Research-Flanders (FWO-Vlaanderen) and the Foundation for
Fundamental Research on Matter (FOM) of the Netherlands. A.Yu. K. thanks NWO 
for financial support.


  \begin{figure}[hbt]
\begin{center}
\leavevmode
\epsfxsize=8cm
{\epsfbox[115 225 535 800]{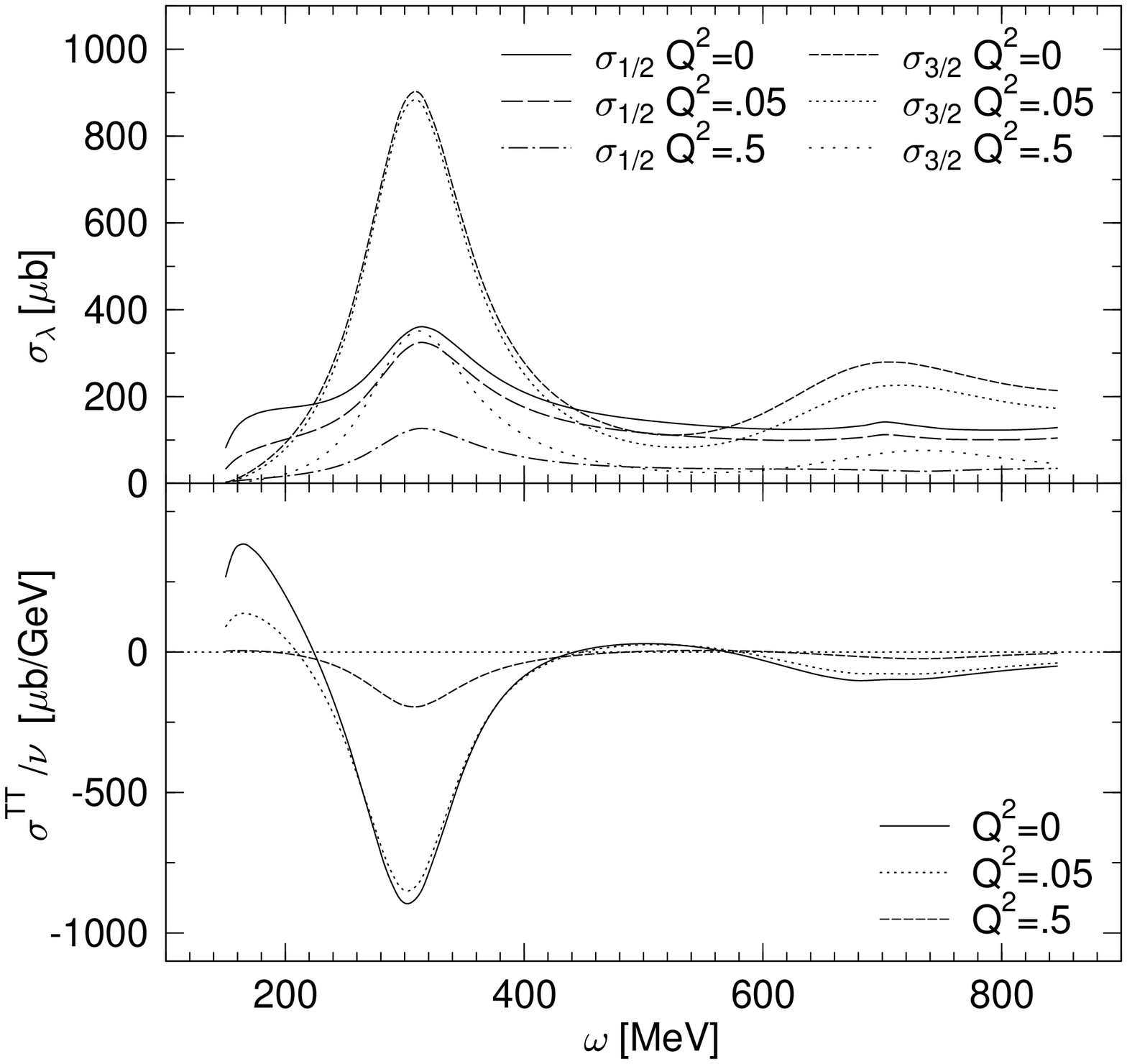}}
\caption[figdhg]{
Energy dependence of the photo-absorption cross section
for parallel ($\sigma_{\thalf}^T$) and anti-parallel ($\sigma_{\half}^T$)
photon and nucleon helicities
at different momentum transfer $Q^2$ as indicated in the figure. The lower
panel shows the energy dependence of the integrand in \eqref{defin}. $Q^2$
is indicated in GeV$^2$.
\figlab{DHG}}
\end{center}
\end{figure}

\newpage

  \begin{figure}[hbt]
\begin{center}
\leavevmode
\epsfxsize=8cm
{\epsfbox{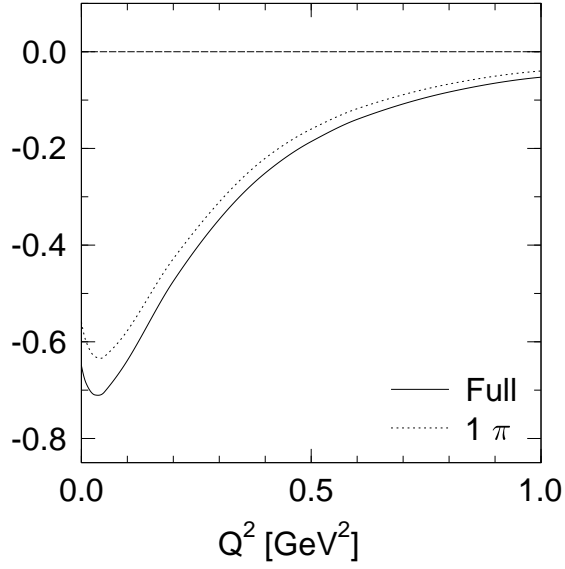}}
\caption[figq]{The momentum dependence of the DHG integral. The dotted line
shows the contribution of the single-pion production channel.
\figlab{DHG-q}}
\end{center}
\end{figure}

  \begin{figure}[hbt]
\begin{center}
\leavevmode
\epsfxsize=8cm
{\epsfbox{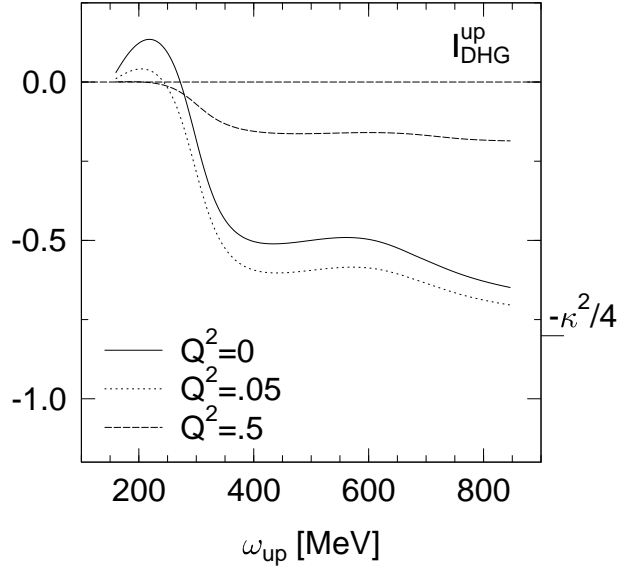}}
\caption[fignu]{Dependence of the integral defined in \eqref{pDHG} on
the upper integration limit for different values of $Q^2$ (in GeV$^2$).
The sumrule value $-\kappa^2/4$ is indicated on the right.
\figlab{DHG-nu}}
\end{center}
\end{figure}

\end{document}